\documentclass[apjl,floatfix,amsmath]{emulateapj}
\usepackage{epsfig,graphicx,graphics,amsfonts,amsmath,amssymb,bm,hyperref,natbib}
\newcommand{\be}{\begin{equation}}
\newcommand{\ee}{\end{equation}}
\newcommand{\bs}{\begin{subequations}}
\newcommand{\es}{\end{subequations}}

\shorttitle{GRAVITATIONAL-WAVE MEMORY REVISITED}
\shortauthors{FAVATA}

\begin{document}
\title{Nonlinear gravitational-wave memory from binary black hole mergers}
\author{Marc Favata}
\address{Kavli Institute for Theoretical Physics, University of California, Santa Barbara, CA 93106-4030, USA}
\submitted{Submitted 2009 February 20; accepted 2009 April 3; published 2009 April 24}
\begin{abstract}
Some astrophysical sources of gravitational waves can produce a ``memory effect,'' which causes a permanent displacement of the test masses in a freely falling gravitational-wave detector. The Christodoulou memory is a particularly interesting nonlinear form of memory that arises from the gravitational-wave stress--energy tensor's contribution to the distant gravitational-wave field. This nonlinear memory contributes a nonoscillatory component to the gravitational-wave signal at leading (Newtonian-quadrupole) order in the waveform amplitude. Previous computations of the memory and its detectability considered only the inspiral phase of binary black hole coalescence. Using an ``effective-one-body'' (EOB) approach calibrated to numerical relativity simulations, as well as a simple fully analytic model, the Christodoulou memory is computed for the inspiral, merger, and ringdown. The memory will be very difficult to detect with ground-based interferometers, but is likely to be observable in supermassive black hole mergers with LISA out to redshifts $z\lesssim 2$. Detection of the nonlinear memory could serve as an experimental test of the ability of gravity to ``gravitate.''
\end{abstract}
\keywords{black hole physics -- gravitation -- gravitational waves -- relativity}
\section{Introduction and motivation}
\begin{figure}[t]
\includegraphics[angle=0, width=0.45\textwidth]{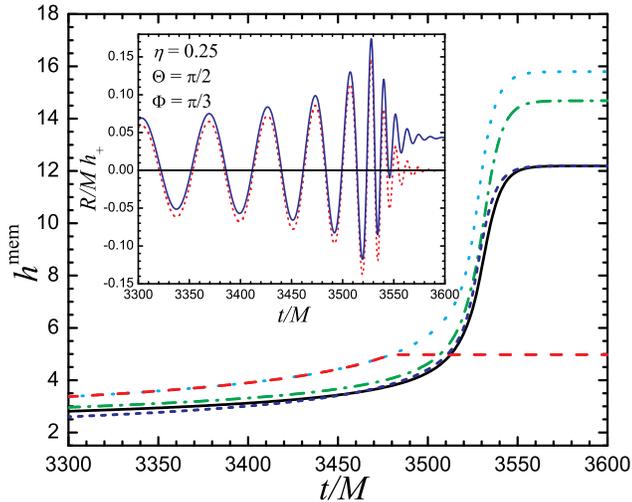}
\caption{\label{fig:hmem}Evolution and saturation of the memory near the merger. The main plot shows several calculations of ${h}^{({\rm mem})}$ [Equation \eqref{eq:hmemhat}]. The solid (black) line uses the full EOB formalism calibrated to numerical relativity simulations (see the text). The dashed-dotted (green) line uses the same formalism but without any EOB amplitude corrections [Equation \eqref{eq:I22EOB} with $r_{\omega} \Omega \rightarrow (M \Omega)^{1/3}$, $F_{22} f_{22}^{\rm NQC}=1$]. The dotted (cyan) curve is the minimal-waveform model [MWM; Equation \eqref{eq:hmemMWM} with $n_{\rm max}=2$ and $r_m=3M$]. The short-dashed (blue) curve is the minimal-waveform model multiplied by a ``fudge factor'' $\approx 0.77$ so that it matches the full-EOB curve at late times. The long-dashed (red) curve is the inspiral memory truncated at $r_m \rightarrow 5M$ [Equation \eqref{eq:hmemMWM} without the sum; this is the model used in Ken94]. All curves (except the ``fudge factor'' one) approach the same value at early times. The inset shows the $h_{+}$ waveform with (blue, solid) and without (red, dashed) memory computed using the full-EOB model [the oscillatory terms contain only the $l=\pm m =2$ modes via Equations \eqref{eq:hdecompose}, \eqref{eq:dnI22ring}, and \eqref{eq:I22EOB}].  All plots are for equal-mass mergers with the matching to the ringdown signal near $t_m/M \approx 3522$.}
\end{figure}

In the typical picture of a coalescing binary black hole (BBH), the gravitational-wave (GW) signal oscillates about a zero value with an amplitude that grows slowly during the inspiral, rises rapidly during merger, and damps back to zero during the ringdown. This picture is not entirely accurate. In actuality, the GW signal displays a \emph{memory}: for a circularized, inspiralling binary there is a growing, nonoscillatory contribution to the``$+$'' polarization that causes the signal amplitude to damp to a nonzero value (see the inset of Figure \ref{fig:hmem}). In a freely falling GW detector, this memory causes a permanent displacement of the test masses that persists after the GW has passed. While at late times the memory yields a constant (undetectable) shift in the spacetime metric, the buildup of the memory is detectable.

For gravitating systems with unbound components, a \emph{linear} memory effect has been known since the 1970s (see the references in Thorne 1992, hereafter Th92): this arises from nearly-zero-frequency changes in the time derivatives of the multipole moments of the source. For example, a hyperbolic binary has linear memory because the multipole moments' time derivatives depend on the (unbound) relative velocity at late and early times, which has the same magnitude but changes direction. Linear memory also appears in systems that undergo a kick (newborn neutron stars, recoiling black holes) or eject particles anisotropically (neutrino emission in supernovae, gamma-ray burst jets) [see Favata 2009b (MF1) and Favata 2009a for references and further discussion].

In the 1990's Blanchet \& Damour (1992, hereafter BD92) and \citet{christodoulou-mem} independently discovered a \emph{nonlinear} memory effect that is present in all GW sources. This ``Christodoulou memory'' arises from the unbound gravitons radiated from the system (Th92): the lost GW energy contributes to the source's changing multipole moments. The Christodoulou memory is a unique manifestation of the nonlinearity of general relativity (GR) and is interesting for several reasons: (1) unlike other nonlinear effects on the GW amplitude such as ``tails'' (backscattering of GWs off of spacetime curvature), the nonlinear memory is nonoscillatory (it builds up over time as GW energy is lost); (2) even more so than tails, the memory is sensitive to the entire past history of the binary's motion \citep{arun25PNamp}; (3) even though the memory originates from the GW stress--energy tensor's contribution at 2.5 post-Newtonian (PN) and higher orders, the memory affects the GW amplitude at \emph{leading} (Newtonian) order\footnote{Partly because of its high PN origin, partly because its nonoscillatory nature makes its detection difficult, and partly because the effect is not widely known, the memory is usually ignored in standard discussions of the leading quadrupole-order waveform.}; and (4) the memory is detectable and its observation could provide a strong-field test of GR.

\begin{figure}[t]
\includegraphics[angle=0, width=0.45\textwidth]{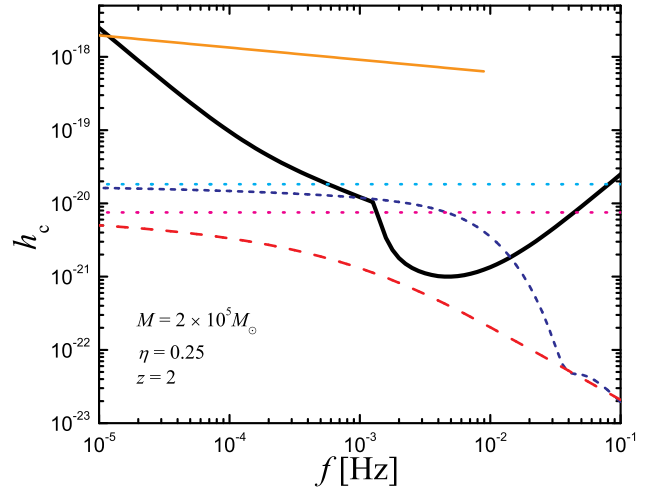}
\caption{\label{fig:hc}Illustration of how the memory model affects the signal-to-noise ratio (SNR). The short-dashed (blue) curve (the main result) shows the characteristic signal strength $h_c$ of the total coalescence memory vs.~frequency [computed using the minimal-waveform model times the factor $0.77$ and Equations \eqref{eq:hc}, \eqref{eq:hplusmem}, and \eqref{eq:FThmemhat}; see the associated time-domain curve in Figure \ref{fig:hmem}]. The SNR corresponding to this model is $8.9$. Contrast this with the long-dashed (red) curve, which includes only the inspiral memory truncated at the last stable orbit (LSO) [see the associated time-domain curve in Figure \ref{fig:hmem}; the Fourier transform in this case is given by Equation \eqref{eq:FThmemhat} without the sum and with $r_m \rightarrow 5M$]. The corresponding SNR is $0.58$. At low frequencies these curves approach the two horizontal (dotted) lines, which represent a signal model that treats the memory as a step function in time, with the size of the step $\Delta h^{\rm (mem)}$ corresponding to the curves in Figure \ref{fig:hmem} that asymptote to $12.2$ or $5.0$.  The corresponding SNRs are $23$ and $9.5$ for the upper and lower dotted horizonal lines. The thick solid (black) curve shows the LISA sky-averaged noise spectrum $h_n(f)$. For comparison the $h_c$ corresponding to the primary (oscillatory) waves is also shown [the top solid (orange) line], truncated at the LSO frequency $f_{\rm LSO} = [5^{3/2} \pi M (1+z)]^{-1}$. The SNR for the inspiral waves is $820$.}
\end{figure}

Previous calculations of the memory for circularized binaries have treated only the inspiral \citep{wiseman-will-memory,blanchet3pnwaveform} and have recently been extended to high PN orders (MF1). Estimates of the memory's detectability have either treated the memory as an unmodeled burst (Th92) or only included the buildup of memory during the inspiral (Kennefick 1994, hereafter Ken94). This work provides the first realistic estimate of the evolution of the memory, accounting for all phases of BBH coalescence (inspiral, merger, ringdown). Properly modeling the memory during the merger and ringdown is especially important because the reciprocal of the memory's rise time [$\sim 1/(70M) \sim 1.45 \text{mHz}\, (2\times 10^6 M_{\odot}/M)$ for a BBH with total mass $M$ and reduced-mass ratio $\eta=0.25$] can lie near the peak sensitivity of ground- and space-based GW detectors, affecting the power at those frequencies. Figures \ref{fig:hmem} and \ref{fig:hc} illustrate the differences between previous memory computations and this work.

Current numerical relativity (NR) simulations can best compute the dominant $l=m=2$ mode of the waveform, as well as other (higher-order) oscillatory modes that contain no physical memory. Because these simulations cannot yet accurately extract the $m=0$ modes that contain memory, a semi-analytic calculation of this effect is necessary. (See MF1 for further discussion.)
\newpage
\section{\label{sec:memcalc}Calculating the memory}
We begin by expanding the GW ``$+$'' and ``$\times$'' polarizations on a basis of spin-weighted spherical harmonics,
\be
\label{eq:hdecompose}
\sqrt{2} R (h_{+} - i h_{\times}) = \sum_{l=2}^{+\infty} \sum_{m=-l}^{l} ( U_{lm} - i V_{lm} ) {}_{-2}Y^{lm}(\Theta,\Phi).
\ee
Here $(R,\Theta,\Phi)$ are the spherical coordinates pointing from the source to the observer; $U_{lm}$ and $V_{lm}$ are the coefficients of the \emph{radiative} mass and current multipole moment tensors ${\mathcal U}_{i_1 i_2 \cdots i_l}$ and ${\mathcal V}_{i_1 i_2 \cdots i_l}$ expanded on the basis of spherical harmonic tensors.\footnote{For $l=2$, $U_{2m}=(16 \pi \sqrt{3}/15) {\mathcal U}_{ij} {\mathcal Y}_{ij}^{2m \ast}$, where the scalar spherical harmonics are $Y^{2m} = {\mathcal Y}_{ij}^{2m} n^i n^j$, $n^i$ is a unit radial vector, and $\ast$ denotes complex conjugation. See MF1 for further notational details.}
The radiative moments are functions of retarded time $T_R$ and appear in the general outgoing solution of the linear vacuum-wave-equation for GWs. The PN wave-generation formalism \citep{blanchetLRRshort} provides a nonlinear, iterative algorithm that relates these radiative moments to a family of \emph{source} multipole moments that are expressible as integrals over the stress--energy pseudotensor of the matter and gravitational fields of the source. This iterative algorithm leads to the following weak-field expansion for the radiative mass multipoles (BD92; when not explicit, $G=c=1$):
\be
\label{eq:Ulmhered}
U_{lm} = I_{lm}^{(l)} + G U_{lm}^{\rm (tail)} + G U_{lm}^{\rm (mem)} + O(G^2) + O(G/c^5).
\ee
Here $I_{lm}^{(l)}$ is the $l^{\rm th}$ time derivative of the source mass moment; $U_{lm}^{\rm (tail)}$ is the GW ``tail'' contribution. The memory piece of the radiative multipole is given by (BD92, MF1)
\be
\label{eq:Ulmmem}
U_{lm}^{\rm (mem)} = \frac{32\pi}{c^{2-l}} \sqrt{\frac{(l-2)!}{2(l+2)!}} \int_{-\infty}^{T_R} \!\! dt \int \! d\Omega \, \frac{dE_{\rm gw}}{dt d\Omega}(\Omega) Y_{lm}^{\ast}(\Omega) ,
\ee
where $\frac{dE_{\rm gw}}{dt d\Omega}$ is the GW energy flux. We ignore the radiative current moments, which have no nonlinear memory.

For an adequate first estimate of the total coalescence memory we focus only on the dominant $l=2$ multipoles. Substituting the leading-order energy flux, $\frac{dE_{\rm gw}}{dt d\Omega} \approx (32\pi)^{-1} \sum_{m,m'} I^{(3)}_{2m} I^{(3)}_{2m'} {}_{-2}Y^{2m} {}_{2}Y^{2m'}$, into Equation \eqref{eq:Ulmmem} and performing the angular integrals gives
the time derivative of the leading-order contribution to the memory piece of the radiative mass moments:
\bs
\label{eq:dUlmleadingorder}
\begin{align}
\label{eq:dU20N}
U_{20}^{{\rm (mem)}(1)} &= \frac{1}{14} \sqrt{\frac{5}{3\pi}} {I}_{22}^{(3)} {I}_{2 -2}^{(3)} [1 + O(c^{-2})], \\
\label{eq:dU40N}
U_{40}^{{\rm (mem)}(1)} &= \frac{1}{2520} \sqrt{\frac{5}{\pi}} {I}_{22}^{(3)} {I}_{2 -2}^{(3)} [1 + O(c^{-2})],
\end{align}
\es
where we have specialized to orbits in the $xy$ plane ($I_{2\pm 1}=0$) and ignored moments that enter at higher PN orders. The $U_{l0}$ for odd-$l$ vanish, and those with $l\geq 6$ enter at higher PN orders. The $U_{lm}^{({\rm mem})}$ for $m\neq 0$ yield oscillatory terms at 2.5PN and higher orders that do not contribute to the memory.

Substituting Equations \eqref{eq:dUlmleadingorder} into Equation \eqref{eq:hdecompose} gives the memory contribution to $h_+$ [$h_{\times}^{\rm (mem)}=0$ for circularized binaries and standard choices of the polarization tensors]:
\begin{align}
\label{eq:hplusmem}
\!\!\! h_{+}^{({\rm mem})} = &\frac{\eta M}{384 \pi R} \sin^2\Theta (17 + \cos^2\Theta) {h}^{({\rm mem})}, \; \text{where} \\
\label{eq:hmemhat}
&{h}^{({\rm mem})} \equiv \frac{1}{\eta M} \int_{-\infty}^{T_R} |{I}_{22}^{(3)}(t)|^2 dt .
\end{align}

To model the evolution of the source-quadrupole moment we follow the ``effective-one-body'' (EOB) approach (see Damour 2008 for references) calibrated to the results of NR simulations. The EOB framework attempts to extend the range of validity of the PN equations of motion to the nonadiabatic ``plunge'' region. It relies on a mapping of the PN 2-body Hamiltonian to the Hamiltonian of a point mass in a \emph{deformed} Schwarzschild metric.

It is instructive to first implement a simple and entirely analytic model for the coalescence that tries to qualitatively capture most of the important physics while minimizing complexity. This \emph{minimal-waveform model} (MWM) consists of matching the leading-order inspiral moments to a sum of quasi-normal modes (QNMs).
During the inspiral the $q^{\rm th}$ derivative of $I_{2\pm2}$ is
\be
\label{eq:dnI22insp}
I_{2\pm 2}^{{\rm insp} (q)} = 2 \sqrt{\frac{2\pi}{5}} \eta M r^2 (\mp 2 i \omega)^q e^{\mp 2 i \varphi},
\ee
where $\omega \equiv \dot{\varphi} = (M/r^3)^{1/2}$ is the orbital frequency, $r=r_m(1-T/\tau_{\rm rr})^{1/4}$ is the orbital separation, $\varphi$ is the 0PN-order orbital phase, $\tau_{\rm rr}=(5/256)(M/\eta)(r_m/M)^4$, $T=t-t_m$, and $r_m$ is the orbital separation at the ``matching time'' $t_m$. For $t>t_m$ the quadrupole-moment derivatives are modeled as a sum of ringdown QNMs:
\be
\label{eq:dnI22ring}
I_{2\pm 2}^{{\rm ring} (2+p)} = \sum_{n=0}^{n_{\rm max}} (-\sigma^{\,}_{22n})^{p} A_{22n} e^{-\sigma^{\,}_{22n}T} ,
\ee
where $\sigma^{\,}_{lmn} = i\omega^{\,}_{lmn} + \tau_{lmn}^{-1}$, with QNM angular frequencies $\omega_{lmn}$ and damping times $\tau_{lmn}$ given by \citet{berti-cardoso-will-PRD2006}. These QNMs depend on the final mass $M_f$ and the dimensionless spin parameter $a_f$ of the BH merger remnant, which are determined by NR simulations (I used the fits to $\eta$ in Table I of \citet{buonanno-pan-baker-etal-nonspinningEOB}). The coefficients $A_{lmn}$ are determined by matching Equations \eqref{eq:dnI22insp}-\eqref{eq:dnI22ring} at $t=t_m$ for $2 \leq (q,p+2) \leq n_{\rm max}+2$.

Substituting these relations into Equation \eqref{eq:hmemhat} (and using $dt=dr/\dot{r}$ for the $t<t_m$ integral) yields
\begin{multline}
\label{eq:hmemMWM}
{h}^{\rm (mem)}_{\rm MWM}(T) = \frac{8\pi M}{r(T)} H(-T) +  H(T) \Bigg\{ \frac{8\pi M}{r_m}   + \frac{1}{\eta M} \\ \!\!\!\! \times \!\!\!\!  \sum_{n,n'=0}^{n_{\rm max}} \!\!\!  \frac{\sigma^{\,}_{22n}\sigma_{22n'}^{\ast} A^{\,}_{22n} A_{22n'}^{\ast}}{\sigma^{\,}_{22n} + \sigma_{22n'}^{\ast}}  \left[ 1 - e^{-(\sigma^{\,}_{22n} + \sigma_{22n'}^{\ast})T} \right] \!\! \Bigg\},
\end{multline}
where $H(T)$ is the Heaviside function.

In addition to this simple, analytic model, I also implement a more realistic EOB description of the moments that closely follows the approach of Damour et al.~(2008b, hereafter DNJ), in which EOB waveforms are calibrated to Jena and Caltech/Cornell NR waveforms:  During the inspiral we solve the EOB equations of motion [\citet{EOB-damour-nagar-finalspin}; Equations (7)-(11)] for $r$, $\varphi$, and the canonical momenta $p_{r^{\ast}}$, and $p_{\varphi}$. For the ``radial potential'' $A(r)$ we use the $(1,4)$-Pad\'{e} resummation of DNJ Equation (6) with $a_5 = 25$. In the $p_{\varphi}$ equation we use DNJ Equations (8)-(11) and Damour \& Nagar (2008, hereafter DNCIT) Equations (17)-(18). For the parameters $\bar{a}_{\rm RR}$ and $v_{\rm pole}$ we use the values in DNJ Table II for $a_5=25$. For initial conditions we use $r_0=15 M$, $\varphi_0=0$, and the ``post-post-circular'' conditions of Damour et al.~(2008a, hereafter DNAEI) Equations (1)-(3) for $(p_{r^{\ast}}, p_{\varphi})$. We replace Equation \eqref{eq:dnI22insp} with
\be
\label{eq:I22EOB}
I_{2\pm2}^{{\rm insp} (q)} = 2 \sqrt{\frac{2\pi}{5}} (\mp 2 i)^q \eta M^{3-q} (r_{\omega} \Omega)^{3q-4} e^{\mp 2 i \varphi} F_{22} f_{22}^{\rm NQC},
\ee
where $r_{\omega}=r \psi^{1/3}$, $\psi = $ [DNJ Equation (11)], $\Omega\equiv \dot{\varphi}$, $F_{22}$ is given in DNCIT Equations (5)-(11), and $f_{22}^{\rm NQC} = $ [DNJ Equation (12)] (with $b=0$ and $a$ given by the linear fit in DNJ Sec.~III). For the ringdown we use Equation \eqref{eq:dnI22ring} above with five QNMs (with modes chosen as in DNAEI, Section III). To determine the coefficients $A_{22n}$ we match $I_{22}^{(2)}$ at five points centered around the time when the radius equals the EOB light-ring [the peak of $\Omega(t)$] as described in DNJ (to improve the fit I shift the matching time by $-3M$). Once constructed $I_{22}^{(3)}$ is substituted into Equation \eqref{eq:hmemhat} and numerically integrated using the initial value ${h}^{\rm (mem)}(0)= 8\pi M/r_0$. The results of the EOB and the minimal-waveform models are shown in Figure \ref{fig:hmem}. Note that the EOB corrections reduce the memory's magnitude.

\section{\label{sec:SNR}Signal-to-noise ratios for the memory}
To compute the memory's signal-to-noise ratio (SNR) we use the MWM [Equation \eqref{eq:hmemMWM}] multiplied by a ``fudge factor'' $\approx 0.77$ that ensures late-time agreement with the full-EOB model (see Figure \ref{fig:hmem}). The sky-averaged $\text{SNR}^2$ for a detector with single-sided noise spectral density $S_n(f)$ is
\be
\label{eq:SNR}
\langle {\rm SNR}^2 \rangle = \int_0^{\infty} \frac{h_c^2(f)}{h_n^2(f)} \frac{df}{f}.
\ee
Here $h_n(f)=\sqrt{\alpha f S_n(f)}$ is the sky-averaged rms noise amplitude, where $\alpha=5$ for orthogonal arm detectors like LIGO and $\alpha=20/3$ for LISA \citep{barackcutler1}. The memory's characteristic amplitude is defined by
\be
\label{eq:hc}
h_c(f) = 2 (1+z) f \langle |\tilde{h}^{\rm (mem)}_{+}[(1+z)f]|^2 \rangle^{1/2}|_{R \rightarrow D_L/(1+z)},
\ee
where $\tilde{h}^{\rm (mem)}_{+}(f)$ denotes the Fourier transform (FT) of Equation \eqref{eq:hplusmem},
$D_L(z)$ is the luminosity distance, and we use the cosmological parameters $H_0/(100\text{ km/s/Mpc}) \approx 0.70$, $\Omega_k=0$, $\Omega_M \approx 0.28$, and $\Omega_{\Lambda} \approx 0.72$.

To compute the FT we use: (1) the FT of $H(T)$, ${\mathcal F}[H(\pm T)] = [\delta(f) \pm i/(\pi f)]/2$; (2) the FT of $H(T) e^{-\beta T}$ for $\beta>0$; and (3) Equation (13.2.6) of \citet{abramowitz-stegun} for the FT of the first term of Equation \eqref{eq:hmemMWM}:
\be
\int_{-\infty}^{0} \frac{e^{2\pi i f T}}{(1-T/\tau_{\rm rr})^{1/4}} dT = \tau_{\rm rr} U(1,7/4, 2\pi i f \tau_{\rm rr}),
\ee
where $U$ is Kummer's confluent hypergeometric function of the second kind.
For $f>0$ the FT of Equation \eqref{eq:hmemMWM} is then
\begin{multline}
\label{eq:FThmemhat}
\!\!\! \tilde{h}^{\rm (mem)}_{\rm MWM}(f) = \frac{i}{2\pi f} \Bigg\{ \! \frac{8\pi M}{r_m} \left[ 1- 2\pi i f \tau_{\rm rr} U(1,7/4,2\pi i f \tau_{\rm rr}) \right] \\   - \frac{1}{\eta M}  \sum_{n,n'=0}^{n_{\rm max}} \frac{\sigma_{22n}^{\,} \sigma_{22n'}^{\ast} A_{22n}^{\,} A_{22n'}^{\ast}}{2\pi i f - (\sigma_{22n}^{\,} + \sigma_{22n'}^{\ast})} \Bigg\}.
\end{multline}

\begin{figure}
\includegraphics[angle=0, width=0.45\textwidth]{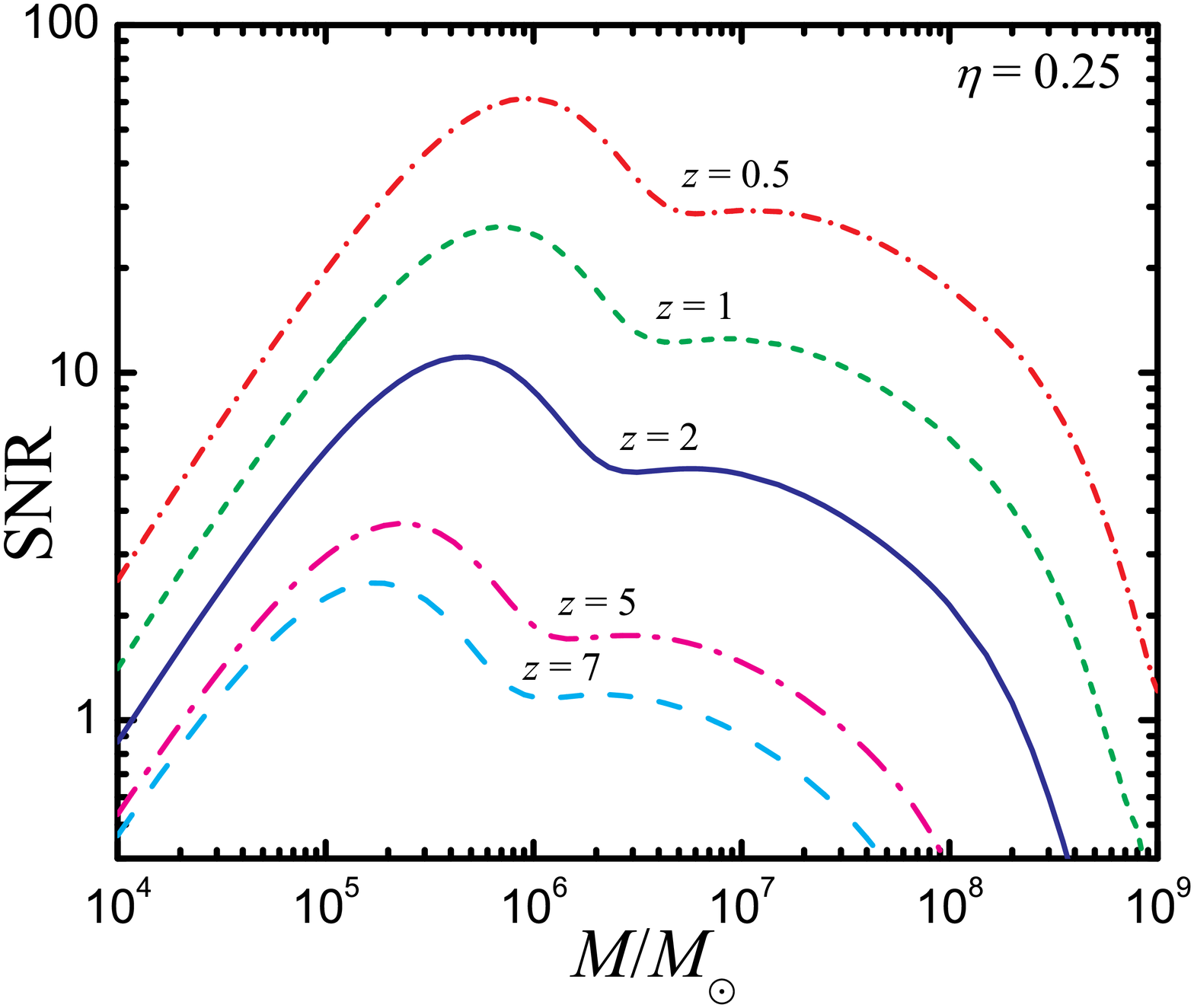}
\caption{\label{fig:SNR}Angle-averaged signal-to-noise ratio of the memory for equal-mass LISA binaries as a function of total binary (source-frame) mass $M$ and redshift $z$. A low-frequency cutoff of $10^{-5}$ Hz in Equation \eqref{eq:SNR} was assumed.}
\end{figure}

Combining the above formulae with sensitivity curves for the various GW interferometers allows us to estimate the memory's detectability. Unless the source is within the Local Group, the memory signal will be too weak to be detected with current LIGO. Advanced LIGO will have a ten times greater sensitivity, yielding a $\text{SNR}\approx 8$ for a $50 M_{\odot}/50 M_{\odot}$ BBH at $20$ Mpc. For LISA the prospects of detecting the memory from supermassive BBH mergers are much better: a $10^5 M_{\odot}/10^5 M_{\odot}$ merger at $z=2$ has a SNR for the memory of $\approx 8.9$ (see Figure \ref{fig:hc}). Figure \ref{fig:SNR} shows the SNR as a function of mass for equal-mass LISA binaries at selected redshifts.

Figure \ref{fig:hc} also illustrates the sensitivity of the SNR to the memory model. Ignoring the merger and ringdown significantly underestimates the memory's SNR. Modeling the memory as a step function in time over-estimates the memory's SNR, even if the correct saturation value $\Delta h^{\rm (mem)}$ is used. This approximation is implicit in the SNR estimate of Th92 and is equivalent to using the zero-frequency-limit [$\tilde{h}^{\rm (mem)}_{\rm ZFL}(f) \approx i \Delta h^{\rm (mem)}/(2\pi f)$] to approximate the memory's FT \citep{smarr-zfl}.
\section{\label{sec:conclusions}Discussion and Conclusions}
The Christodoulou memory is a dramatic example of how the nonlinearities of general relativity can manifest themselves in a detectable GW signal. This first attempt at including the merger and ringdown illustrates the importance of the full coalescence in any model of the memory. This calculation also indicates that despite the recent successes of numerical relativity (NR), analytic methods can still be useful in understanding the highly nonlinear regime of BH mergers. It is especially interesting to see that a simple analytic description like the ``minimal-waveform model'' can qualitatively describe the memory. This is partly due to the memory's independence of the GW phase, which is more sensitive to PN corrections than the amplitude [see Equation \eqref{eq:hmemhat}].

Although based on an EOB model calibrated to NR simulations, these memory estimates relied on various approximations.  While the memory is not easily extracted from current NR simulations (MF1), input from NR is needed to compute the memory accurately. Preliminary results from a hybrid PN/NR calculation using the $l=m=2$ Caltech/Cornell merger waveform \citep{scheel-merger} indicate that the memory saturates at a value $\Delta h^{\rm (mem)} \approx 9.6$ (M.~Favata 2009, in preparation). This $\approx 27\%$ difference with the full-EOB model used here [$\Delta h^{\rm (mem)} \approx 12.2$; see Figure \ref{fig:hmem}] is roughly consistent with the $\approx 20\%$ error between the amplitude of the NR waveform and the EOB model of DNJ near merger (see their Figures 8 and 9). Future work will consider higher multipole interactions and recent improvements in the EOB formalism \citep{damour-iyer-nagar}.

Signal-to-noise ratio (SNR) calculations suggest that the memory from stellar-mass BH mergers is unlikely to be detected with Advanced LIGO, but the memory from supermassive BH mergers is potentially detectable by LISA out to a redshift $z\lesssim 2$ (Figure \ref{fig:SNR}).
This detectability assessment is based on the memory's SNR lying above some threshold $\sim 5$--$8$. A better measure of detectability should consider separating the memory from the stronger oscillatory merger and ringdown waves. This will be addressed in later work. Future Mock LISA Data Challenges \citep{mldc-cqg07} should consider incorporating the memory into their data sets. The memory also introduces new parameter dependencies into the GW signal, and it would be interesting to see how this affects the estimation of binary parameters \citep{lisaparametertaskforce-lisa7}. If the nonlinear memory is eventually detected, it could provide an interesting test of general relativity, particularly of the ability of gravity to ``gravitate.''
\acknowledgements
This research was supported in part by the National Science Foundation under Grant No.~PHY05-51164. I thank Emanuele Berti, Alessandra Buonanno, Scott Hughes, and the referee for helpful comments on this manuscript.

\end{document}